\documentclass[prl, twocolumn, amsmath, amssymb,showpacs,hyperref,superscriptaddress,showpacs]{revtex4}
\usepackage{graphicx}
\usepackage{dcolumn}
\usepackage{bm}
\usepackage{xspace}
\usepackage{color}
\usepackage{epstopdf}
\usepackage[
  bookmarks%
  ,pdftitle={Search for Supersymmetry with Like-Sign Lepton-Tau Events at CDF}%
  ,pdfsubject={CDF Run 2 physics result}%
  ,pdfauthor={CDF Collaboration}%
  ,pdfstartview={Fit}%
  ]{hyperref}
\begin{document}

%
\newcommand{\capt}[3]{ \begin{minipage}{0.8\textwidth} \caption[#1]{
      \renewcommand{\baselinestretch}{1} \small #2} \label{#3} \end{minipage}  } 
\newcommand{\beq}{\begin{equation}}
\newcommand{\eeq}{\end{equation}}
\newcommand{\bbf}{\begin{bfseries}}
\newcommand{\ebf}{\end{bfseries}}
\newcommand{\beqn}{\begin{eqnarray}}
\newcommand{\eeqn}{\end{eqnarray}}
\newcommand{\bi}{\begin{itemize}}
\newcommand{\ei}{\end{itemize}}
\newcommand{\bd}{\begin{description}}
\newcommand{\ed}{\end{description}}
\newcommand{\ben}{\begin{enumerate}}
\newcommand{\een}{\end{enumerate}}
\newcommand{\bHuge}{\begin{Huge}}
\newcommand{\bhuge}{\begin{huge}}
\newcommand{\bLARGE}{\begin{LARGE}}
\newcommand{\bLarge}{\begin{Large}}
\newcommand{\blarge}{\begin{large}}
\newcommand{\eHuge}{\end{Huge}}
\newcommand{\ehuge}{\end{huge}}
\newcommand{\eLARGE}{\end{LARGE}}
\newcommand{\eLarge}{\end{Large}}
\newcommand{\elarge}{\end{large}}
\newcommand{\blue}{\color{blue}}
\newcommand{\red}{\color{red}}
\newcommand{\redd}[1]{ \textcolor{red}{#1} }
\newcommand{\blued}[1]{ \textcolor{blue}{#1} }
\newcommand{\magd}[1]{ \textcolor{magenta}{#1} }
\newcommand{\greend}[1]{ \textcolor{mydgreen}{#1} }
\newcommand{\blackd}[1]{ \textcolor{black}{#1} }
\newcommand{\black}{\color{black}}
\newcommand{\green}{\color{mydgreen}}
\newcommand{\magenta}{\color{magenta}}
\newcommand{\mybull}{\textcolor{magenta}{$\lozenge$}}
\newcommand{\loz}{\item [\mybull]}
%
%
%
%
\newcommand{\pbar}{\mbox{$\overline{p}$}}
\newcommand{\ppbar}{\mbox{$p\overline{p}$}}
\newcommand{\bbbar}{\mbox{$b\overline{b}$}}
\newcommand{\ccbar}{\mbox{$c\overline{c}$}}
\newcommand{\ttbar}{\mbox{$t\overline{t}$}}
\newcommand{\wpm}{\mbox{$W^{\pm}$}}
\newcommand{\zzero}{\mbox{$Z^0$}}
\newcommand{\ztt}{\mbox{$Z^0\to\tau\tau$}}
\newcommand{\ee}{\mbox{$ee$}}
\newcommand{\emu}{\mbox{$e\mu$}}
\newcommand{\mumu}{\mbox{$\mu\mu$}}
\newcommand{\etal}{{\it et al.}}
\newcommand{\met}{\mbox{${E\!\!\!\!/_T}$}}
\newcommand{\rpv}{\mbox{${R\!\!\!\!\!\:/_p}$}}
\newcommand{\rp}{\mbox{${R_p}$}}
\newcommand{\lnv}{\mbox{${L\!\!\!\!\!\:/}$}}
\newcommand{\bnv}{\mbox{${B\!\!\!\!\!\:/}$}}
\newcommand{\lamp}{\mbox{$\lambda_{121}'$}}
\newcommand{\intlum}{\mbox{${ \int {\cal L} \; dt}$}}
\newcommand{\pt}{\mbox{$p_T$}}
\newcommand{\ipt}{\mbox{$p^{-1}_{{\rm T}}$}}
\newcommand{\ptcut}{\mbox{$p_{{\rm T}}$-Cut}}
\newcommand{\et}{\mbox{$E_T$}}
\newcommand{\sigpt}{\mbox{$\sigma_{\pt}/\pt$}}
\newcommand{\dphi}{\mbox{$\Delta\varphi$}}
\newcommand{\lxy}{\mbox{$L_{xy}$}}
%
%
\newcommand{\pelp}{\mbox{$e^+$}}
\newcommand{\pelm}{\mbox{$e^-$}}
\newcommand{\pelpm}{\mbox{$e^{\pm}$}}
\newcommand{\epem}{\mbox{$e^+e^-$}}
\newcommand{\mpmm}{\mbox{$\mu^+\mu^-$}}
\newcommand{\plp}{\mbox{$l^+$}}
\newcommand{\plm}{\mbox{$l^-$}}
\newcommand{\pmup}{\mbox{$\mu^+$}}
\newcommand{\pmum}{\mbox{$\mu^-$}}
\newcommand{\pprp}{\mbox{$\rm{p}^+$}}
\newcommand{\pprm}{\mbox{$\rm{p}^-$}}
\newcommand{\pprb}{\mbox{${\overline{p}}$}}
\newcommand{ \lpm    }{\mbox{$\ell^{\pm}$}}
\newcommand{ \epm    }{\mbox{$e^{\pm}$}}
\newcommand{ \lplus  }{\mbox{$\ell^+$}}
\newcommand{ \lminus }{\mbox{$\ell^-$}}
\newcommand{ \lplm   }{\mbox{$\ell^+\ell^-$}}
\newcommand{\Pkao}{\mbox{$\mathrm{K}^0$}}
\newcommand{\Pkab}{\mbox{$\mathrm\overline{K}^0$}}
\newcommand{\kao}{\mbox{$K^0$}}
\newcommand{\kab}{\mbox{$\overline{K}^0$}}
%
%
\newcommand{ \qbar}   {\mbox{$\overline{q}$}}
\newcommand{ \photino}{\mbox{$\tilde{\gamma}$}}
\newcommand{ \gluino} {\mbox{$\tilde{g}$}}
\newcommand{ \glgl}   {\mbox{$\tilde{g}\tilde{g}$}}
\newcommand{ \squark} {\mbox{$\tilde{q}$}}
\newcommand{ \sqbar } {\mbox{$\bar{\tilde{q}}$}}
\newcommand{ \mgluino} {\mbox{$M(\gluino)$}}
\newcommand{ \msquark} {\mbox{$M(\squark)$}}
\newcommand{ \csquarkl} {\mbox{$\tilde{c}_L$}}
\newcommand{ \magcs} {\textcolor{magenta}{\csquarkl}}
\newcommand{ \mcsl} {\mbox{$M(\csquarkl)$}}
\newcommand{ \bcsquarkl} {\mbox{$\bar{\tilde{c}}_L$}}
\newcommand{ \magcsb} {\textcolor{magenta}{\bcsquarkl}}
\newcommand{ \cscsb}   {\mbox{$\csquarkl\bcsquarkl$}}
\newcommand{ \ssb}     {\mbox{$\squark\overline{\squark}$}}
\newcommand{ \usquark} {\mbox{$\tilde{u}$}}
\newcommand{ \uub}     {\mbox{$\usquark\overline{\usquark}$}}
\newcommand{ \dsquark} {\mbox{$\tilde{d}$}}
\newcommand{ \csquark} {\mbox{$\tilde{c}$}}
\newcommand{ \ssquark} {\mbox{$\tilde{s}$}}
\newcommand{ \tsquark} {\mbox{$\tilde{t}$}}
\newcommand{ \stopo}   {\mbox{$\tilde{t}_1$}}
\newcommand{ \ttb}     {\mbox{$\tsquark\overline{\tsquark}$}}
\newcommand{ \ttbone}  {\mbox{$\tsquark_1\overline{\tsquark}_1$}}
\newcommand{ \bsquark} {\mbox{$\tilde{b}$}}
\newcommand{ \sneu   } {\mbox{$\tilde{\nu}$}}
\newcommand{ \sele   } {\mbox{$\tilde{e}$}}
\newcommand{ \stau   } {\mbox{$\tilde{\tau}$}}
\newcommand{ \slep   } {\mbox{$\tilde{\ell}$}}
\newcommand{ \wino}   {\mbox{$\tilde{W}^{\pm}$}}
\newcommand{ \zino}   {\mbox{$\tilde{Z}$}}
\newcommand{\sfermion}{\mbox{$\widetilde f$}}
\newcommand{ \chizero}{\mbox{$\tilde{\chi}_{1}^0$}}
\newcommand{ \none}{\mbox{$\tilde{\chi}_{1}^0$}}
\newcommand{ \lsp} {\mbox{$\tilde{\chi}_{1}^0$}}
\newcommand{ \ntwo}{\mbox{$\tilde{\chi}_{2}^0$}}
\newcommand{ \nthree}{\mbox{$\tilde{\chi}_{3}^0$}}
\newcommand{ \nfour}{\mbox{$\tilde{\chi}_{4}^0$}}
\newcommand{ \mchio}{\mbox{$M(\chizero)$}}
\newcommand{ \mnone}{\mbox{$M(\chizero)$}}
\newcommand{ \chione }{\mbox{$\tilde{\chi}_{1}^{\pm}$}}
\newcommand{ \cone }{\mbox{$\tilde{\chi}_{1}^{\pm}$}}
\newcommand{ \ctwo }{\mbox{$\tilde{\chi}_{2}^{\pm}$}}
\newcommand{ \mchione }{\mbox{$M(\tilde{\chi}_{1}^{\pm})$}}
\newcommand{ \mcone }{\mbox{$M(\tilde{\chi}_{1}^{\pm})$}}
\newcommand{ \mstopo}   {\mbox{$M(\tilde{t}_1)$}}
\newcommand{ \chipm  }{\mbox{$\tilde{\chi}^{\pm}$}}
\newcommand{ \chitwo }{\mbox{$\tilde{\chi}_{2}^0$}}
\newcommand{ \mchitwo }{\mbox{$M(\tilde{\chi}_{2}^0)$}}
\newcommand{ \chichi} {\mbox{$\chione\chitwo$}}
\newcommand{ \chipchim}{\mbox{$\tilde{\chi}_{1}^+\tilde{\chi}_{1}^-$}}
\newcommand{ \sqmass }{\mbox{$M_{\tilde{q}}$}}
\newcommand{ \glmass }{\mbox{$M_{\tilde{g}}$}}
\newcommand{ \mz     }{\mbox{$M_0$}}
\newcommand{ \mo     }{\mbox{$M_{1/2}$}}
%
\newcommand{\mll}{\mbox{$M(\ell\ell)$}}
\newcommand{\mjj}{\mbox{$M(jj)$}}
\newcommand{\mlljj}{\mbox{$M(\ell\ell jj)$}}
\newcommand{\dphill}{\mbox{$\Delta \phi (\ell\ell)$}}
\newcommand{\dphijj}{\mbox{$\Delta \phi (jj)$}}
%

%
%
\newcommand{\dagg}[1]{ #1 ^{\dagger}}
\newcommand{\modulus}[1]{\left| #1 \right|}
\newcommand{\paren}[1]{\left( #1 \right)}
\newcommand{\ave}[1]{\left\langle #1 \right\rangle}
\newcommand{\mods}[1]{\left| #1 \right|^2}
\newcommand{\bra}[1]{\langle #1 |}
\newcommand{\ket}[1]{| #1 \rangle}
\newcommand{\braket}[2]{\langle #1 | #2 \rangle}
\newcommand{\bracket}[3]{\langle #1 | #2 | #3 \rangle}
\newcommand{\sqbs}[1]{\left[ #1 \right] }
\newcommand{\rea}[1]{\mbox{$\,{\rm Re}\left( #1 \right)$}}
\newcommand{\ima}[1]{\mbox{$\,{\rm Im}\left( #1 \right)$}}
%
%
%
\newcommand{\mev}  {\mbox{$\;{\rm MeV}/c$}}
\newcommand{\mevc} {\mbox{$\;{\rm MeV}/c^2$}}
\newcommand{\gev}  {\mbox{$\;{\rm GeV}$}}
\newcommand{\tev}  {\mbox{$\;{\rm TeV}$}}
\newcommand{\gevc} {\mbox{$\;{\rm GeV}/c$}}
\newcommand{\gevcc}{\mbox{$\;{\rm GeV}/c^2$}}
\newcommand{\tevcc}{\mbox{$\;{\rm TeV}/c^2$}}
\newcommand{\cmtwo}{\mbox{${\rm cm}^2$}}
\newcommand{\mmtwo}{\mbox{${\rm mm}^2$}}
\newcommand{\lumin}{\mbox{${\rm cm^{-2}\,s^{-1}}$}}
%
%
\newcommand{\ipb}{\mbox{${\rm pb}^{-1}$}}
\newcommand{\ifb}{\mbox{${\rm fb}^{-1}$}}
%
%
\newcommand{\chis}{\mbox{$\chi^{2}$}}
\newcommand{\ie}{i.e.}
\newcommand{\eg}{e.g.}
\newcommand{\dedx}{\mbox{${\rm d}E/{\rm d}x$}}
\newcommand{\dndx}{\mbox{${\rm d}n/{\rm d}x$}}
\newcommand{\micron}{\mbox{$\mu{\rm m}$}}
\newcommand{\musec}{\mbox{$\mu {\rm s}$}}
\newcommand{\eps}{\mbox{$\epsilon$}}
\newcommand{\dxdy}{\mbox{${\rm d}X{\rm d}Y$}}
\newcommand{\sigrphi}{\mbox{$\sigma_{r \phi}$}}
\newcommand{\sigz}{\mbox{$\sigma_{z}$}}
\newcommand{\deltapp}{\mbox{$\Delta p/p$}} 
\newcommand{\dzero}{\mbox{${\rm D\O}$}}
\newcommand{\svx}{SVX II}
%
%
%
%
%

\affiliation{Institute of Physics, Academia Sinica, Taipei, Taiwan 11529, Republic of China}
\affiliation{Argonne National Laboratory, Argonne, Illinois 60439, USA}
\affiliation{University of Athens, 157 71 Athens, Greece}
\affiliation{Institut de Fisica d'Altes Energies, ICREA, Universitat Autonoma de Barcelona, E-08193, Bellaterra (Barcelona), Spain}
\affiliation{Baylor University, Waco, Texas 76798, USA}
\affiliation{Istituto Nazionale di Fisica Nucleare Bologna, $^{ee}$University of Bologna, I-40127 Bologna, Italy}
\affiliation{University of California, Davis, Davis, California 95616, USA}
\affiliation{University of California, Los Angeles, Los Angeles, California 90024, USA}
\affiliation{Instituto de Fisica de Cantabria, CSIC-University of Cantabria, 39005 Santander, Spain}
\affiliation{Carnegie Mellon University, Pittsburgh, Pennsylvania 15213, USA}
\affiliation{Enrico Fermi Institute, University of Chicago, Chicago, Illinois 60637, USA}
\affiliation{Comenius University, 842 48 Bratislava, Slovakia; Institute of Experimental Physics, 040 01 Kosice, Slovakia}
\affiliation{Joint Institute for Nuclear Research, RU-141980 Dubna, Russia}
\affiliation{Duke University, Durham, North Carolina 27708, USA}
\affiliation{Fermi National Accelerator Laboratory, Batavia, Illinois 60510, USA}
\affiliation{University of Florida, Gainesville, Florida 32611, USA}
\affiliation{Laboratori Nazionali di Frascati, Istituto Nazionale di Fisica Nucleare, I-00044 Frascati, Italy}
\affiliation{University of Geneva, CH-1211 Geneva 4, Switzerland}
\affiliation{Glasgow University, Glasgow G12 8QQ, United Kingdom}
\affiliation{Harvard University, Cambridge, Massachusetts 02138, USA}
\affiliation{Division of High Energy Physics, Department of Physics, University of Helsinki and Helsinki Institute of Physics, FIN-00014, Helsinki, Finland}
\affiliation{University of Illinois, Urbana, Illinois 61801, USA}
\affiliation{The Johns Hopkins University, Baltimore, Maryland 21218, USA}
\affiliation{Institut f\"{u}r Experimentelle Kernphysik, Karlsruhe Institute of Technology, D-76131 Karlsruhe, Germany}
\affiliation{Center for High Energy Physics: Kyungpook National University, Daegu 702-701, Korea; Seoul National University, Seoul 151-742, Korea; Sungkyunkwan University, Suwon 440-746, Korea; Korea Institute of Science and Technology Information, Daejeon 305-806, Korea; Chonnam National University, Gwangju 500-757, Korea; Chonbuk National University, Jeonju 561-756, Korea; Ewha Womans University, Seoul, 120-750, Korea}
\affiliation{Ernest Orlando Lawrence Berkeley National Laboratory, Berkeley, California 94720, USA}
\affiliation{University of Liverpool, Liverpool L69 7ZE, United Kingdom}
\affiliation{University College London, London WC1E 6BT, United Kingdom}
\affiliation{Centro de Investigaciones Energeticas Medioambientales y Tecnologicas, E-28040 Madrid, Spain}
\affiliation{Massachusetts Institute of Technology, Cambridge, Massachusetts 02139, USA}
\affiliation{Institute of Particle Physics: McGill University, Montr\'{e}al, Qu\'{e}bec H3A~2T8, Canada; Simon Fraser University, Burnaby, British Columbia V5A~1S6, Canada; University of Toronto, Toronto, Ontario M5S~1A7, Canada; and TRIUMF, Vancouver, British Columbia V6T~2A3, Canada}
\affiliation{University of Michigan, Ann Arbor, Michigan 48109, USA}
\affiliation{Michigan State University, East Lansing, Michigan 48824, USA}
\affiliation{Institution for Theoretical and Experimental Physics, ITEP, Moscow 117259, Russia}
\affiliation{University of New Mexico, Albuquerque, New Mexico 87131, USA}
\affiliation{The Ohio State University, Columbus, Ohio 43210, USA}
\affiliation{Okayama University, Okayama 700-8530, Japan}
\affiliation{Osaka City University, Osaka 588, Japan}
\affiliation{University of Oxford, Oxford OX1 3RH, United Kingdom}
\affiliation{Istituto Nazionale di Fisica Nucleare, Sezione di Padova-Trento, $^{ff}$University of Padova, I-35131 Padova, Italy}
\affiliation{University of Pennsylvania, Philadelphia, Pennsylvania 19104, USA}
\affiliation{Istituto Nazionale di Fisica Nucleare Pisa, $^{gg}$University of Pisa, $^{hh}$University of Siena and $^{ii}$Scuola Normale Superiore, I-56127 Pisa, Italy, $^{mm}$INFN Pavia and University of Pavia, I-27100 Pavia, Italy}
\affiliation{University of Pittsburgh, Pittsburgh, Pennsylvania 15260, USA}
\affiliation{Purdue University, West Lafayette, Indiana 47907, USA}
\affiliation{University of Rochester, Rochester, New York 14627, USA}
\affiliation{The Rockefeller University, New York, New York 10065, USA}
\affiliation{Istituto Nazionale di Fisica Nucleare, Sezione di Roma 1, $^{jj}$Sapienza Universit\`{a} di Roma, I-00185 Roma, Italy}
\affiliation{Mitchell Institute for Fundamental Physics and Astronomy, Texas A\&M University, College Station, Texas 77843, USA}
\affiliation{Istituto Nazionale di Fisica Nucleare Trieste/Udine; $^{nn}$University of Trieste, I-34127 Trieste, Italy; $^{kk}$University of Udine, I-33100 Udine, Italy}
\affiliation{University of Tsukuba, Tsukuba, Ibaraki 305, Japan}
\affiliation{Tufts University, Medford, Massachusetts 02155, USA}
\affiliation{University of Virginia, Charlottesville, Virginia 22906, USA}
\affiliation{Waseda University, Tokyo 169, Japan}
\affiliation{Wayne State University, Detroit, Michigan 48201, USA}
\affiliation{University of Wisconsin, Madison, Wisconsin 53706, USA}
\affiliation{Yale University, New Haven, Connecticut 06520, USA}

\author{T.~Aaltonen}
\affiliation{Division of High Energy Physics, Department of Physics, University of Helsinki and Helsinki Institute of Physics, FIN-00014, Helsinki, Finland}
\author{S.~Amerio}
\affiliation{Istituto Nazionale di Fisica Nucleare, Sezione di Padova-Trento, $^{ff}$University of Padova, I-35131 Padova, Italy}
\author{D.~Amidei}
\affiliation{University of Michigan, Ann Arbor, Michigan 48109, USA}
\author{A.~Anastassov$^x$}
\affiliation{Fermi National Accelerator Laboratory, Batavia, Illinois 60510, USA}
\author{A.~Annovi}
\affiliation{Laboratori Nazionali di Frascati, Istituto Nazionale di Fisica Nucleare, I-00044 Frascati, Italy}
\author{J.~Antos}
\affiliation{Comenius University, 842 48 Bratislava, Slovakia; Institute of Experimental Physics, 040 01 Kosice, Slovakia}
\author{G.~Apollinari}
\affiliation{Fermi National Accelerator Laboratory, Batavia, Illinois 60510, USA}
\author{J.A.~Appel}
\affiliation{Fermi National Accelerator Laboratory, Batavia, Illinois 60510, USA}
\author{T.~Arisawa}
\affiliation{Waseda University, Tokyo 169, Japan}
\author{A.~Artikov}
\affiliation{Joint Institute for Nuclear Research, RU-141980 Dubna, Russia}
% ...

\collaboration{CDF Collaboration\footnote{With visitors from
$^a$University of British Columbia, Vancouver, BC V6T 1Z1, Canada,
$^b$Istituto Nazionale di Fisica Nucleare, Sezione di Cagliari, 09042 Monserrato (Cagliari), Italy,
$^c$University of California Irvine, Irvine, CA 92697, USA,
$^e$Institute of Physics, Academy of Sciences of the Czech Republic, 182~21, Czech Republic,
$^f$CERN, CH-1211 Geneva, Switzerland,
$^g$Cornell University, Ithaca, NY 14853, USA,
$^{dd}$The University of Jordan, Amman 11942, Jordan,
$^h$University of Cyprus, Nicosia CY-1678, Cyprus,
$^i$Office of Science, U.S. Department of Energy, Washington, DC 20585, USA,
$^j$University College Dublin, Dublin 4, Ireland,
$^k$ETH, 8092 Z\"{u}rich, Switzerland,
$^l$University of Fukui, Fukui City, Fukui Prefecture, Japan 910-0017,
$^m$Universidad Iberoamericana, Lomas de Santa Fe, M\'{e}xico, C.P. 01219, Distrito Federal,
$^n$University of Iowa, Iowa City, IA 52242, USA,
$^o$Kinki University, Higashi-Osaka City, Japan 577-8502,
$^p$Kansas State University, Manhattan, KS 66506, USA,
$^q$Brookhaven National Laboratory, Upton, NY 11973, USA,
$^r$University of Manchester, Manchester M13 9PL, United Kingdom,
$^s$Queen Mary, University of London, London, E1 4NS, United Kingdom,
$^t$University of Melbourne, Victoria 3010, Australia,
$^u$Muons, Inc., Batavia, IL 60510, USA,
$^v$Nagasaki Institute of Applied Science, Nagasaki 851-0193, Japan,
$^w$National Research Nuclear University, Moscow 115409, Russia,
$^x$Northwestern University, Evanston, IL 60208, USA,
$^y$University of Notre Dame, Notre Dame, IN 46556, USA,
$^z$Universidad de Oviedo, E-33007 Oviedo, Spain,
$^{aa}$CNRS-IN2P3, Paris, F-75205 France,
$^{cc}$Universidad Tecnica Federico Santa Maria, 110v Valparaiso, Chile,
$^{ll}$Universite catholique de Louvain, 1348 Louvain-La-Neuve, Belgium,
$^{oo}$University of Z\"{u}rich, 8006 Z\"{u}rich, Switzerland,
$^{pp}$Massachusetts General Hospital and Harvard Medical School, Boston, MA 02114 USA,
$^{qq}$Hampton University, Hampton, VA 23668, USA,
$^{rr}$Los Alamos National Laboratory, Los Alamos, NM 87544, USA
}}
\noaffiliation

\thispagestyle{empty}
\vspace*{-2cm}
\title
  {\bf Search for Supersymmetry with Like-Sign Lepton-Tau Events at CDF}
\author{The CDF Collaboration}
\affiliation{URL \url{http://www-cdf.fnal.gov}}

\begin{abstract}
We present a search for chargino-neutralino associated production using like electric charge dilepton events collected by the CDF II detector at the Fermilab Tevatron in proton-antiproton collisions at $\sqrt{s} = 1.96$ TeV.    One lepton is identified as the hadronic decay of a tau lepton, while the other is an electron or muon.  
In data corresponding to 6.0 fb$^{-1}$ of integrated luminosity, we obtain good agreement with standard model predictions, and set limits on the chargino-neutralino production cross section for simplified gravity- and gauge-mediated models.  
As an example, assuming that the chargino and neutralino decays to taus dominate, in the simplified gauge-mediated model
we exclude cross sections greater than 300 fb at 95\% credibility level for chargino and neutralino masses of 225 \gevcc.
This analysis is the first to extend the LHC searches for electroweak supersymmetric production of gauginos to high $\tan\beta$ and slepton next-to-lightest supersymmetric particle scenarios.
\end{abstract}


\maketitle

Supersymmetry (SUSY) is an appealing extension to the standard model (SM)
of particle physics as it mitigates the hierarchy problem, provides a dark matter
candidate, and allows for gauge-coupling unification at high energy 
~\cite{ref:SUSY-1, ref:SUSY0, ref:SUSY1,
  ref:SUSY2, ref:SUSY3, ref:SUSY4, ref:hierarchy1, ref:hierarchy2}.
Extensive searches for SUSY phenomena have been performed at the  
  LEP~\cite{LEPLimits}, Tevatron~\cite{Abulencia:2007rd,CDFtrileptons,Aaltonen:2007af,
  CDFLimits, D0Limits, Abazov200934},
  and LHC~\cite{PhysRevLett.106.131802,
  Aad2011186, springerlink:10.1140/epjc/s10052-011-1682-6,
RA1Paper, CMSmultilepton1,CMSmultilepton2} colliders.  To date,
no evidence of SUSY has been found.  The LHC analyses provide
stringent limits on the SUSY partners of light quarks and the gluon, 
the squarks and the gluino, with mass limits in excess of 1 $\tevcc$.
Typical searches assume strong production of squarks and gluinos with cascade
decays to the gauginos (the SUSY partners of the electroweak gauge and Higgs bosons, the charginos and neutralinos), followed by hadronic or leptonic 
decays.  These final-state particles are accompanied by two or more of the lightest SUSY particle
(LSP), that is stable if $R_p$ parity is conserved~\cite{Farrar:1978xj}.  In the minimal supersymmetric
standard model (MSSM) with gravity mediation, the LSP is often the lightest neutralino \none, which provides a cosmological dark matter candidate.  Alternatively, in gauge-mediated models~\cite{Dimopoulos:1996vz,Giudice:1998bp}, the gravitino plays the role of the LSP, and the phenomenology depends on the nature of the next-to-lightest SUSY particles.  If these are the SUSY lepton partners (sleptons), their decays lead to detectable leptons.  Both models produce an appreciable momentum imbalance in the plane transverse to the beam direction due to the undetected LSPs~\cite{Ruderman:2010kj}.

Given the lack of evidence of strongly-produced SUSY particles, searches for direct electroweak production of charginos and neutralinos are particularly well-motivated at present.  This production can lead to the striking signature of sparse events with two or three leptons and a transverse momentum imbalance.  
Most SUSY searches also assume $\tan\beta \lesssim 10$, where $\tan\beta$ is the ratio of the vacuum expectation
values for the two Higgs doublets, which results in similar gaugino decay-widths to electrons,
muons, and tau leptons.  At high values of $\tan\beta$, e.g., $\tan\beta \simeq 30$, appreciable left-right
mixing drives the mass of the lighter SUSY tau particle (stau, $\stau$) to lower values, and results in enhanced branching fractions to taus as two-body decays become kinematically accessible.  
As the value of $\tan\beta$ is a free parameter of the theory, searches
sensitive to tau leptons can play a critical role in the search for SUSY phenomena.  
ATLAS~\cite{atlasewkino} and CMS~\cite{cmsewkino} have recently published searches for SUSY electroweak production with leptonic decays.  ATLAS searches for trilepton signals with electrons and muons in the final state, and does not consider tau-enriched scenarios.  CMS searches for dilepton and trilepton signals including those with hadronic tau decays, and places bounds on flavor-universal and tau-enriched scenarios.  While these results are generally more stringent than what is possible at the Tevatron, there are regions of parameter space still unexplored by the LHC experiments.  These include the high $\tan\beta$ case where all gaugino decays produce taus, and gauge-mediated scenarios with slepton next-to-lightest SUSY particles.  The current situation provides strong motivation for this analysis, which probes these unexplored regions for the first time.  

This Letter reports the results of a search for chargino-neutralino (\cone\ntwo) associated electroweak production yielding tau-dominated final states using data collected with the CDF II detector at the Fermilab Tevatron $p\bar{p}$ collider at a center-of-mass energy of 1.96 TeV.  The analysis considers a single $W$-boson-mediated $s$-channel amplitude, while the $t$-channel squark exchange amplitude is insignificant with the assumption of heavy squarks, as motivated by the LHC limits.  Using a simplified framework~\cite{Ruderman:2010kj}, we study two distinct cases.  In the first, charginos decay promptly into a
single lepton through a slepton $\tilde{\chi}_1^{\pm}  \rightarrow ~ \tilde{\ell}^{\pm(*)} ~\nu_{\ell} \rightarrow
\tilde{\chi}_1^{0} ~{\ell}^{\pm} ~\nu_{\ell} $ and neutralinos similarly decay into two detectable
leptons $\tilde{\chi}_2^{0}  \rightarrow ~ \tilde{\ell}^{\pm(*)} ~{\ell}^{\mp}
\rightarrow \tilde{\chi}_1^{0} ~{\ell}^{\pm} ~{\ell}^{\mp}. $  The second case assumes the same gaugino decays, followed by the gauge-mediated slepton decays $ \tilde{\ell} \to \ell \tilde{G}$, where $\tilde{G}$ is the LSP gravitino.
Both cases yield events with three electrically-charged leptons accompanied by undetectable particles.  However, requiring the detection of all three leptons would degrade the search sensitivity, especially for the case of decays to tau leptons, which is the focus of this analysis.  Instead, we require detection of either an electron or muon plus a hadronically-decaying tau lepton.   Tau leptons decay hadronically, with a branching fraction of about 65\%, as $\tau \to X_h \,\nu_{\tau}$, where $X_h$ is a system of hadrons consisting of charged and neutral pions or kaons. 
A like-sign (LS) requirement on the light lepton ($e,\mu$) electric charge and net electric charge of the tau decay-products efficiently rejects prominent SM backgrounds such as $Z$ boson, $WW$ bosons, and top-antitop quark production, which yield opposite-sign (OS) leptons.  
We perform a counting experiment and compare the yield of LS lepton-tau events in data with SM background predictions folded with sources of misidentified taus, and validate the results with control samples of OS events.  In this Letter, ``lepton" and ``tau" (or $\tau$) refer to $e$ or $\mu$ and hadronically-decaying tau leptons, respectively.  The LS signature is common in many SUSY models.  Our search has sensitivity for high $\tan \beta$ due to a dedicated tau reconstruction, and since the identified $e$ or $\mu$ can result from a leptonic tau decay.

The CDF II detector is described in Ref.~\cite{Abulencia:2005ix}.  The innermost components are multi-layer silicon-strip detectors and an open-cell 
drift chamber tracking system covering $|\eta| < 1$ \cite{cdfgeometry}
inside a 1.4 T superconducting solenoid.  Surrounding the
magnet are sampling electromagnetic and hadronic calorimeters, segmented in
projective-tower geometry, covering $|\eta| < 3.6$.
Strip-wire chambers in the central electromagnetic calorimeter at a depth approximately corresponding to the maximum development of the typical electromagnetic shower aid in reconstructing electrons, photons, 
and $\pi^0 \to \gamma\gamma$ decays in the region $|\eta| < 1.1$.  At larger radii are 
scintillators and wire-chambers for muon identification: the central muon ($|\eta| < 0.6$)
and the forward muon ($0.6 < |\eta| < 1$) detectors.

Data corresponding to an integrated luminosity of 6.0 fb$^{-1}$, collected between 2002 and 2010 by a dedicated online event-selection (trigger) \cite{Anastassov:2003vc},  are used.  This trigger requires a charged particle reconstructed with the silicon and drift chamber detectors with $\pt > 8 \gevc$
matched to an electron (muon) signal in the central electromagnetic calorimeter 
(central or forward muon detector),
and an additional isolated charged particle with $\pt > 5 \gevc$ that seeds the tau reconstruction.  
At trigger level a charged particle is isolated if no additional charged particles with $p_T > 1.5 \gevc$ 
are reconstructed in the annular region between 10 and 30 degrees around the track direction.
No requirement on the relative charge of the lepton and tau is imposed at the trigger level, providing a control sample.  

The total trigger efficiency is the product of the efficiency for selecting a tau and the efficiency for selecting a lepton.
These are determined using independent data samples of  multijet and high-\pt\ lepton events
\cite{Abulencia:2007iy,Abulencia:2005ix}. Jets are sprays of hadronic particles produced in the fragmentation and hadronization of quarks and gluons, and are clustered using a fixed-cone
algorithm \cite{Bhatti:2005ai} with a radius $\Delta R = \sqrt{(\Delta{\eta})^2 + (\Delta{\phi})^2} = 0.4$.  Jets with $\et > 8 \gev$ and $|\eta| < 2.5$ are used.
Here, $\Delta \eta$ ($\Delta \phi$) is the difference relative to the jet axis in $\eta$ ($\phi$) space.
Comparison with simulated $Z\to\tau\tau$ events yields a trigger efficiency for real
taus inside the detector-acceptance region of $(91 \pm 3)\%$ \cite{Abulencia:2007iy}. 
The trigger efficiencies for  reconstructed electrons, central muons, and forward muons are $(96.0 \pm 0.3)\%$,  $(86.6 \pm 0.7)\%$, and $(89.9 \pm 0.7)\%$, respectively \cite{Abulencia:2005ix}.  
These efficiencies include a degradation by less than $10\%$ with increasing number of 
overlapping \ppbar\ interactions per bunch crossing that occur at high-luminosity Tevatron operations.

The event selection proceeds as follows.
Electrons  (muons) are required to satisfy an \et\ (\pt) requirement of 10\gev\ (\gevc), along with quality criteria to increase the purity of the samples ~\cite{Abulencia:2005ix}.  In particular, electrons and muons must be isolated in the tracker and calorimeters, satisfying $\Sigma \pt^{\mathit iso} < 2.0 \gevc$ and $E^{\mathit iso}/ \et < 0.1$ or $E^{\mathit iso} < 2.0 \gev$.
Here $\Sigma \pt^{\mathit iso}$ is the sum of the transverse momenta of any additional charged particles
in a cone of radius $\Delta R  = 0.4$ 
around the candidate lepton, and $E^{\mathit iso}$ is the additional energy deposited in the calorimeters in the same cone.
Hadronic tau decays are identified as systems of one (``one-prong") or three (``three-prong") charged particles in a narrow cone, pointing toward a central calorimeter cluster with $|\eta|<1$.
Momenta of photons from neutral pions are reconstructed using the central
shower-maximum detector. The visible transverse energy of the tau candidate, 
defined as $\pt_{\tau} = \Sigma \pt_{\mathit tracks} + \Sigma\et_{\pi^0}$, must be greater
than 15 (20) \gevc\ for one-prong (three-prong) taus.  Upper thresholds on the tau
invariant mass and calorimeter or tracker activity in an
isolation annulus built around the highest \pt\ (leading) track reduce contamination from quark and gluon jets.
Additional criteria on the ratio of deposited calorimeter energy to leading track \pt\ reject electrons and muons that could mimic the signal \cite{Abulencia:2005kq}.

The event energy-imbalance transverse to the beam direction ($\vec{\met}$) is defined by 
$\vec{\met} = - \sum_{i} E_T^i \hat{n}_i$, where the sum is over all calorimeter towers with
$|\eta| < 3.6$ and $\hat{n}_i$ is a unit vector perpendicular to the beam axis and pointing at the $i$th calorimeter tower.   We also define $\met=|\vec{\met}|$.  To reduce the considerable backgrounds from the production of multijet events,
we use a requirement on the scalar sum ($H_T$) of  \pt\ of the tau, \pt\ of the lepton, and \met.
We require  $H_T > 45 \gev$ (50 \gev) for one-prong tau plus muon (electron) events, and 
$H_T > 55 \gev$ for events with three-prong taus \cite{Forrest:2011zz}.
We require $\Delta \phi (\ell,\tau) > 0.5$ to ensure that the lepton and tau isolation cones do not overlap, and remove events with OS same-flavor leptons consistent with $Z$ boson decay.


Depending on the relative charges of the lepton and the tau, events that pass
the selection are divided into an OS control region and an LS signal region.
The OS control region is mainly composed of SM processes yielding real taus, such as Drell-Yan, \ttbar, and diboson production, plus events with jets misidentified as taus.  These large backgrounds would overwhelm any potential SUSY signal.  For the LS signal region, events with misidentified jets are dominant; these include events with a $W$ boson produced in association with jets ($W$ + jets),  multijet production, and events with photon conversions to $e^+e^-$ pairs.  Because of the kinematic similarity between the SUSY signal and $W$ + jet events, the latter dominates the background composition.  Backgrounds from lepton or tau charge mismeasurement are insignificant \cite{Abulencia:2005ix}.

Backgrounds are estimated using a combination of Monte Carlo (MC) simulations and data-driven methods.  The most significant backgrounds after the LS requirement are due to jet misidentification and are determined directly from data.
We use the \textsc{pythia 6} MC simulation \cite{Sjostrand:2007gs} to generate samples of
events that produce genuine taus from diboson, \ttbar\, and $Z$ boson processes, while $W\to\tau\nu$
events are generated using \textsc{alpgen 2.10$'$} \cite{Alpgen2002} interfaced with \textsc{pythia}
for parton showering and hadronization.  These samples
are processed with the CDF II detector simulation based on \textsc{geant 3} \cite{Agostinelli:2002hh}.  
The sample sizes are normalized to their SM cross sections \cite{Beringer:1900zz} and are appropriately scaled to account for MC-data differences in trigger, identification, and reconstruction efficiencies.

The jet-to-tau misidentification rate is determined using jet-triggered events in data
to account for the dominant background processes, extending the treatment in Refs.~\cite{Abulencia:2005kq,Aaltonen:2009vf}.  
As quark jets and gluon jets are misidentified as taus with different probabilities, we apply a correction for gluon-jet dominated $\gamma$ + jets events with $\gamma \to e^+e^-$ \cite{Forrest:2011zz}.   We parameterize the misidentification rates in terms of $\eta_{\tau}$, the number of tracks in the tau signal cone, 
and the total \et\ in the tau signal and isolation cones, and apply these rates to jets in events that satisfy the remaining selection criteria to determine this contribution to the final event sample.
We verify this technique using data samples enriched in multijet events, selected by
requiring at least 3 \gevc\ (\gev) of additional $\pt$ (\et) in the tracking system (calorimeters).
We also verify this technique in $W$ + jets events, by requiring a $W$-like event topology, and in $\gamma$ + jets events, by requiring $\gamma\to e^+e^-$.

The main source of systematic uncertainty arises from the jet-to-tau misidentification rate,
taken as the misidentification-rate difference between the leading and second-highest-\pt\ jets (25\%).  
These jets are the most likely to be misidentified as taus.  Less significant are uncertainties on the
SM background processes cross-sections (ranging from 2 to 10\%) and the uncertainty on the integrated luminosity (6\%).   The 30\% uncertainty on the photon-conversion-finding efficiency has only a minor effect on the final result.  
We consider a possible systematic uncertainty on the reconstructed tau energy by comparing \pt\ spectra for one- and three-pronged taus in data and simulated $W\to \tau \nu$ samples.  The best agreement is obtained by shifting the tau energy scale in the simulation by 1\%.  Finally, the uncertainty on the hadronic jet-energy scale leads to a 1.5\% systematic uncertainty on the reconstructed tau energy for events with real taus.

\begin{figure}[htbp]
  \begin{center}
    \includegraphics[width=7cm,clip=]{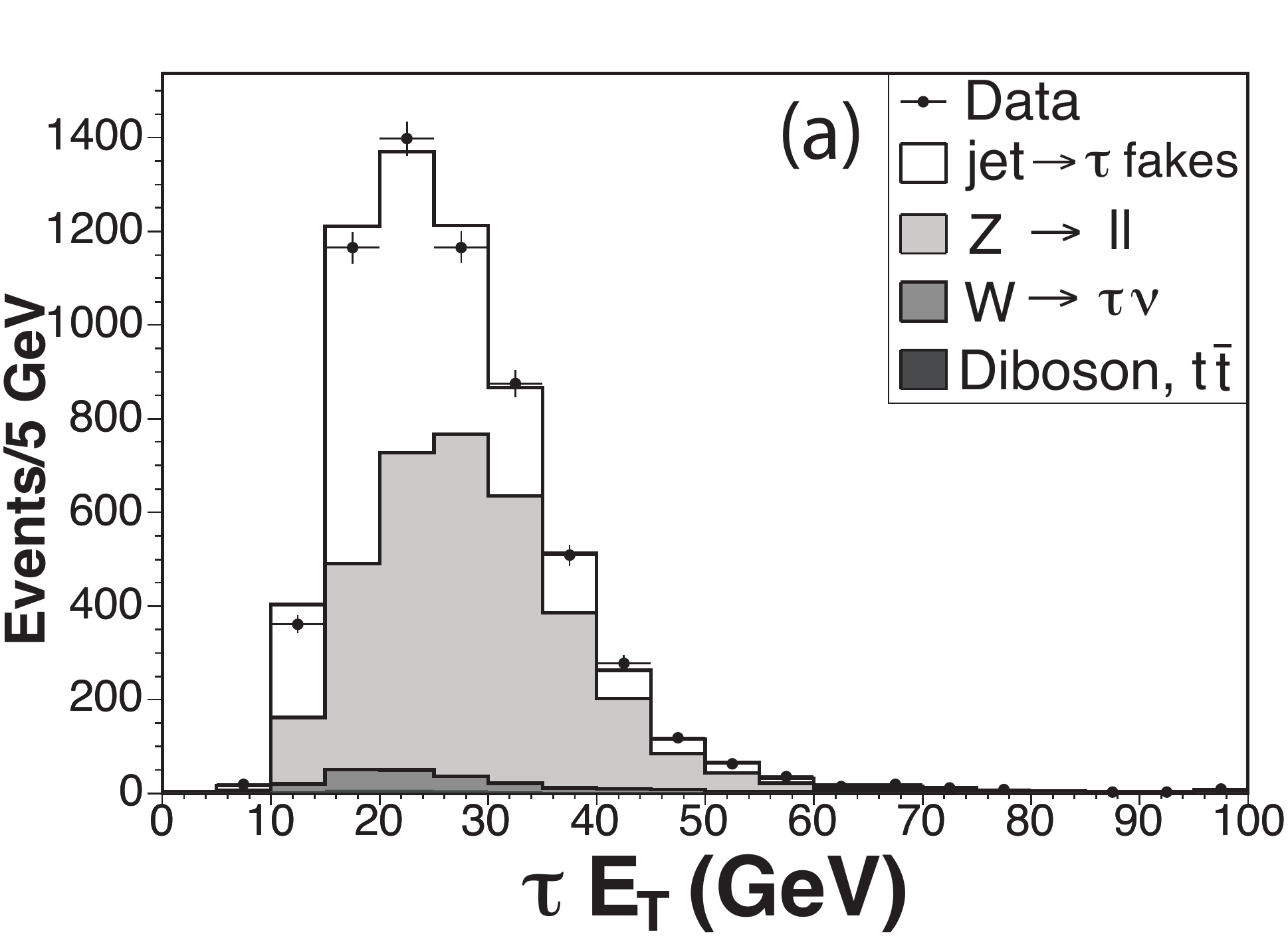}
    \includegraphics[width=7cm,clip=]{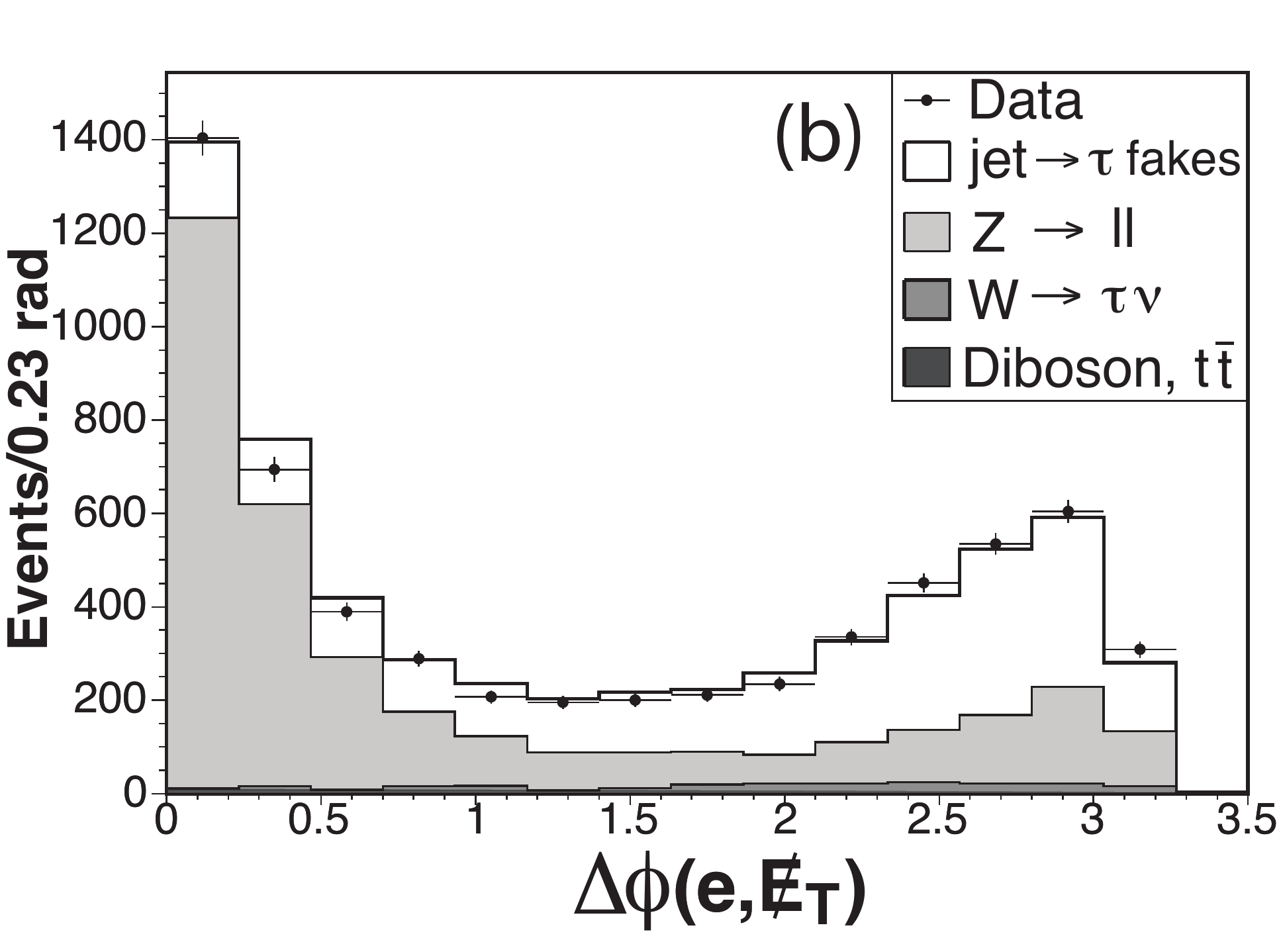}
        \includegraphics[width=7cm,clip=]{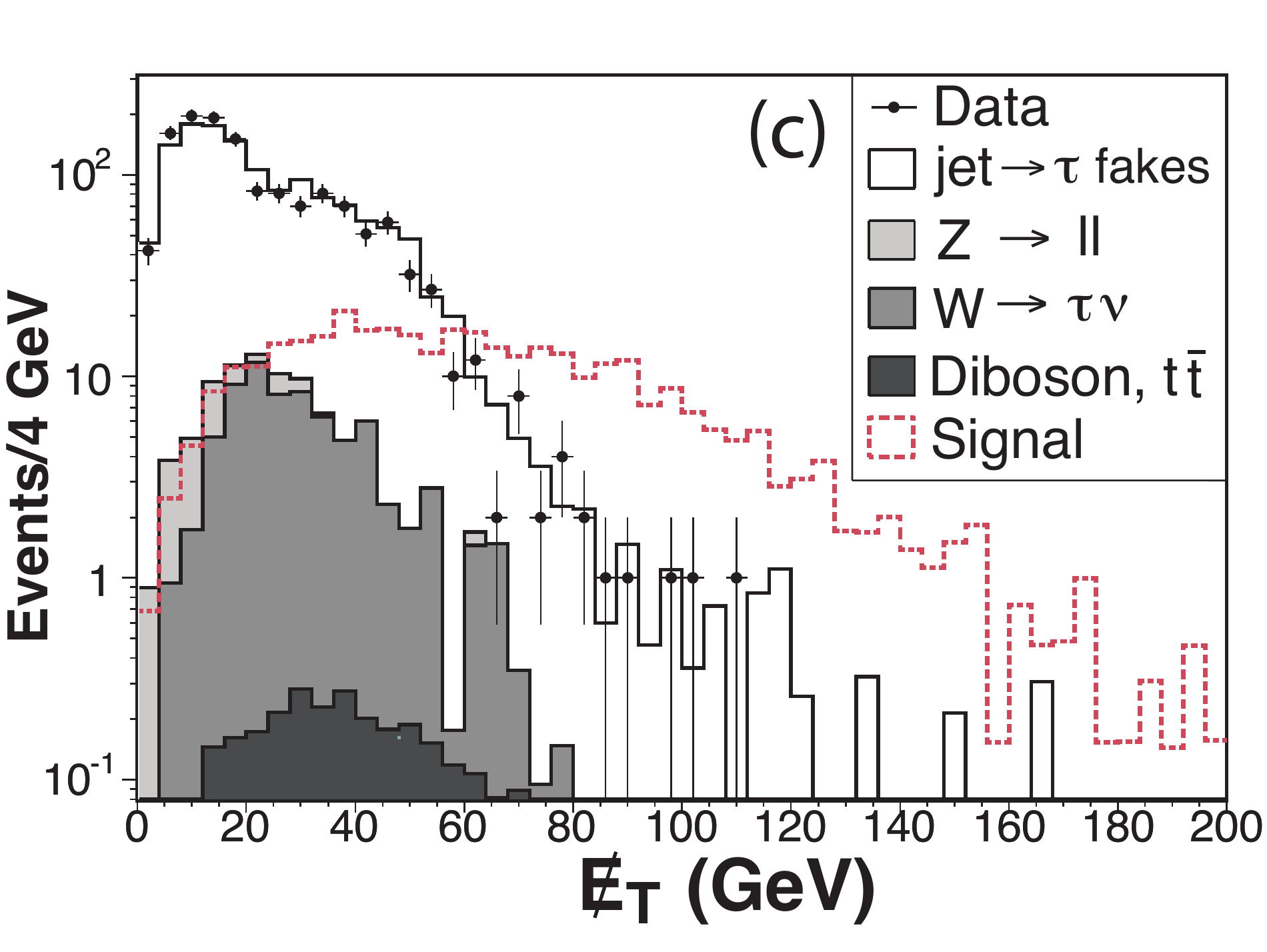}
\caption{Distribution of  (a) tau cluster \et\ and (b) $\Delta \phi(e,\met)$ for OS $e+\tau$ events.  
Distribution of (c) \met\ for LS (signal-region) $\mu+\tau$ events.  Overlaid is a signal distribution corresponding to the simplified gauge-mediated
  model, with $\sigma(\cone\ntwo) = $ 3000 fb for visibility, $m(\cone) = m(\ntwo) =200 \gevcc$, 
  and $m(\tilde{\ell}) = 160 \gevcc$.
\label{fig1}}
  \end{center}
\end{figure}

The background determination is validated using 
the OS control region.  Results are given in Table~\ref{tab:bgdata}, and show good agreement in both the OS control region and in the LS signal region.  Figure \ref{fig1} shows representative kinematic distributions for the OS control region and the LS signal region.    

\begin{table}
 \centering
   \caption{Backgrounds and observations in data for OS control region and LS signal region.  The
  signal region values include the $H_T$ requirement described in the text.  For each entry, the statistical, followed  by the systematic uncertainty, is given.  The signal corresponds to the simplified gauge-mediated model, with $\sigma(\cone\ntwo) = $ 300 fb, $m(\cone) = m(\ntwo) =200 \gevcc$, and $m(\tilde{\ell}) = 160 \gevcc$.  For this specific scenario, the optimized requirement is $\met > 98 \gev$.}
 \mbox{
 \begin{tabular}{lrr}
\hline\hline
Process & OS events  & LS events\\
 & & ($\met>20 \gev$)\\ \hline
$Z\rightarrow \tau \tau $ & $6967\pm 56\pm 557$ & 10$\pm$2$\pm$1 \\
Jet$\rightarrow \tau $ & $4527  \pm 27 \pm 1065$  & 1153$\pm$15$\pm$283\\
$Z\rightarrow \mu \mu $ & $263 \pm  20 \pm 21$  &  \multicolumn{1}{c}{$\--$} \\
$Z\rightarrow e e $ & $83  \pm  9 \pm 7 $               & \multicolumn{1}{c}{$\--$} \\
$W \rightarrow \tau \nu $ &  $372  \pm 12 \pm 36 $  & 97$\pm$6$\pm$10\\
$t\bar{t} $ & $36.3 \pm 0.3 \pm 5.1 $                                    & 0.7$\pm$0.0$\pm$0.1\\
Diboson & $61 \pm 1 \pm 6 $                                      & 4.3$\pm$0.2$\pm$0.4\\ \hline
Total & 12308 $\pm\ 67\pm 1202$                              & 1265$\pm$17$\pm$283\\
Data & 12268 & 1116\\ 
\hline
& Signal & 64 $\pm$ 1 $\pm$ 6 \\
&  Optimized $\met$ requirement & ($\met > 98 \gev$) \\
& Total background& 6 $\pm$ 1 $\pm$ 1 \\
& Signal & 10 $\pm$ 1 $\pm$ 1\\
& Data &3 \\
\hline\hline
\end{tabular} }
\label{tab:bgdata}
  \end{table}

\begin{figure}[htbp]
  \begin{center}
    \includegraphics[width=7cm,clip=]{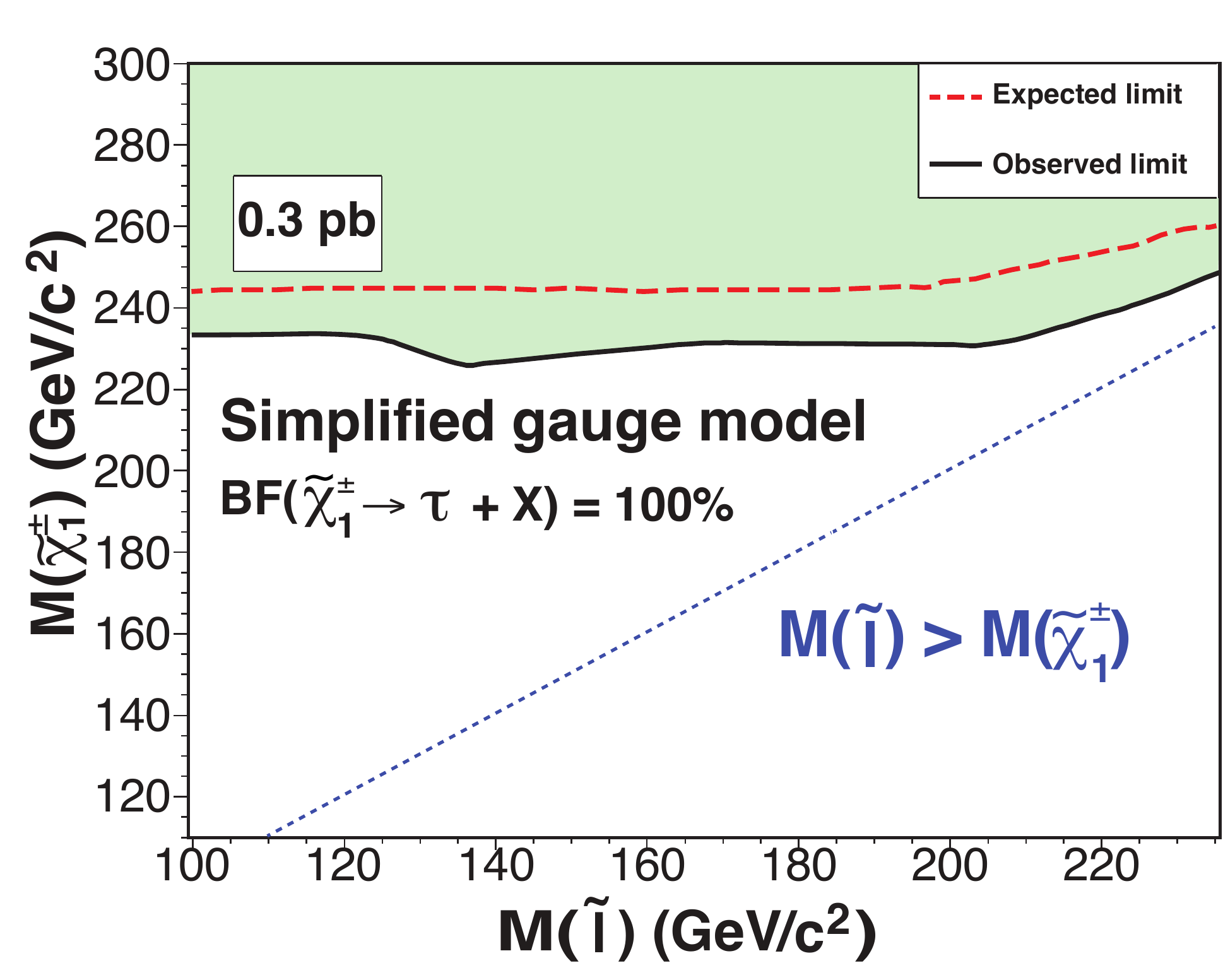}
    \caption{Expected and observed contours of constant 95\% C.L. cross-section upper limit in the chargino-slepton mass plane assuming the simplified gauge-mediated model for ${\it BF}(\tilde{\chi} \to \tau + X)=100\%$.  The shaded region corresponds to cross section limits of $\sigma(\cone\ntwo) \leq $ 300 fb, as a function of the gaugino and slepton masses.
    \label{fig:limits-gauge}}
  \end{center}
\end{figure}

\begin{figure}[htbp]
  \begin{center}
    \includegraphics[width=7cm,clip=]{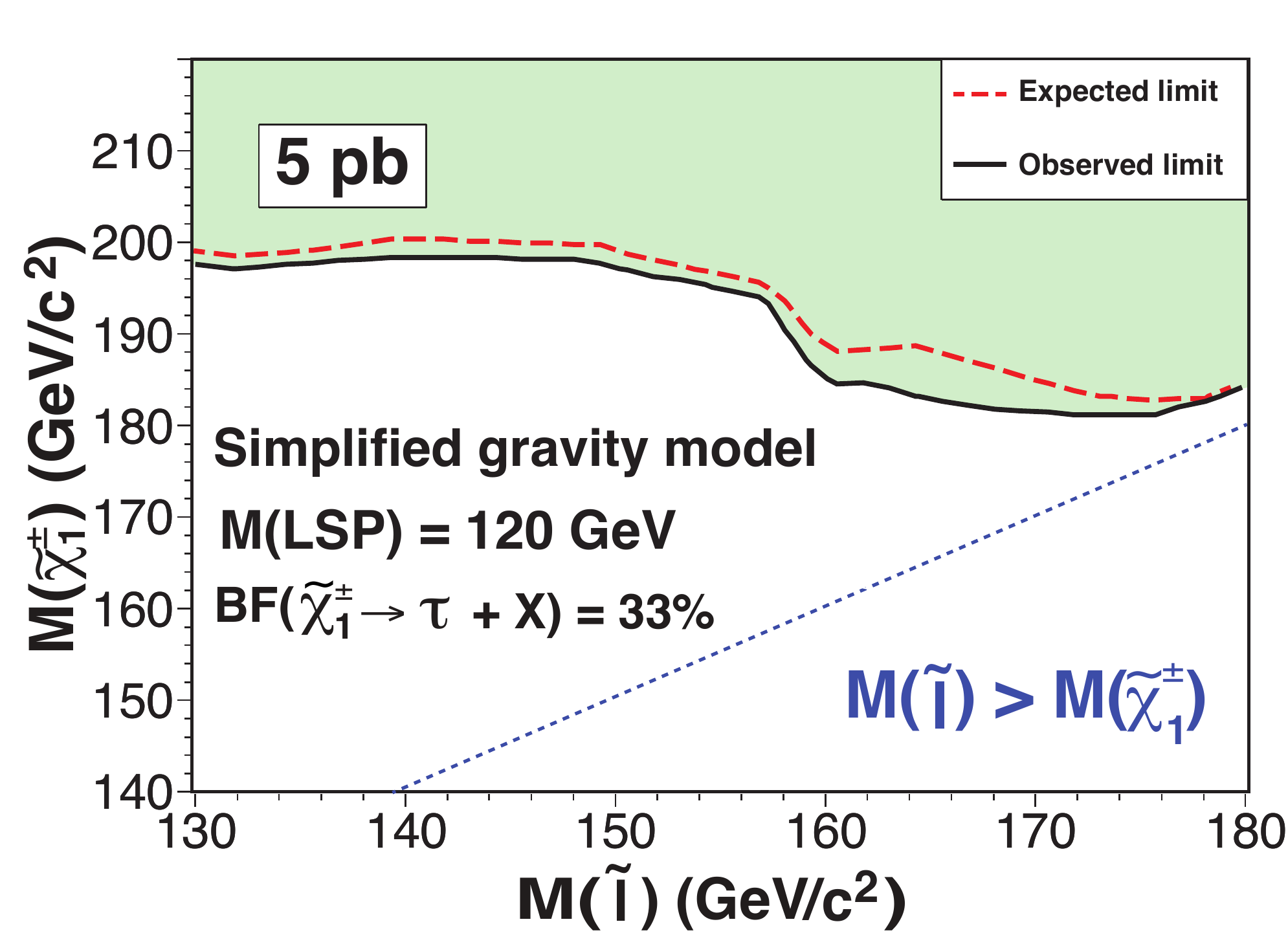}
        \includegraphics[width=7cm,clip=]{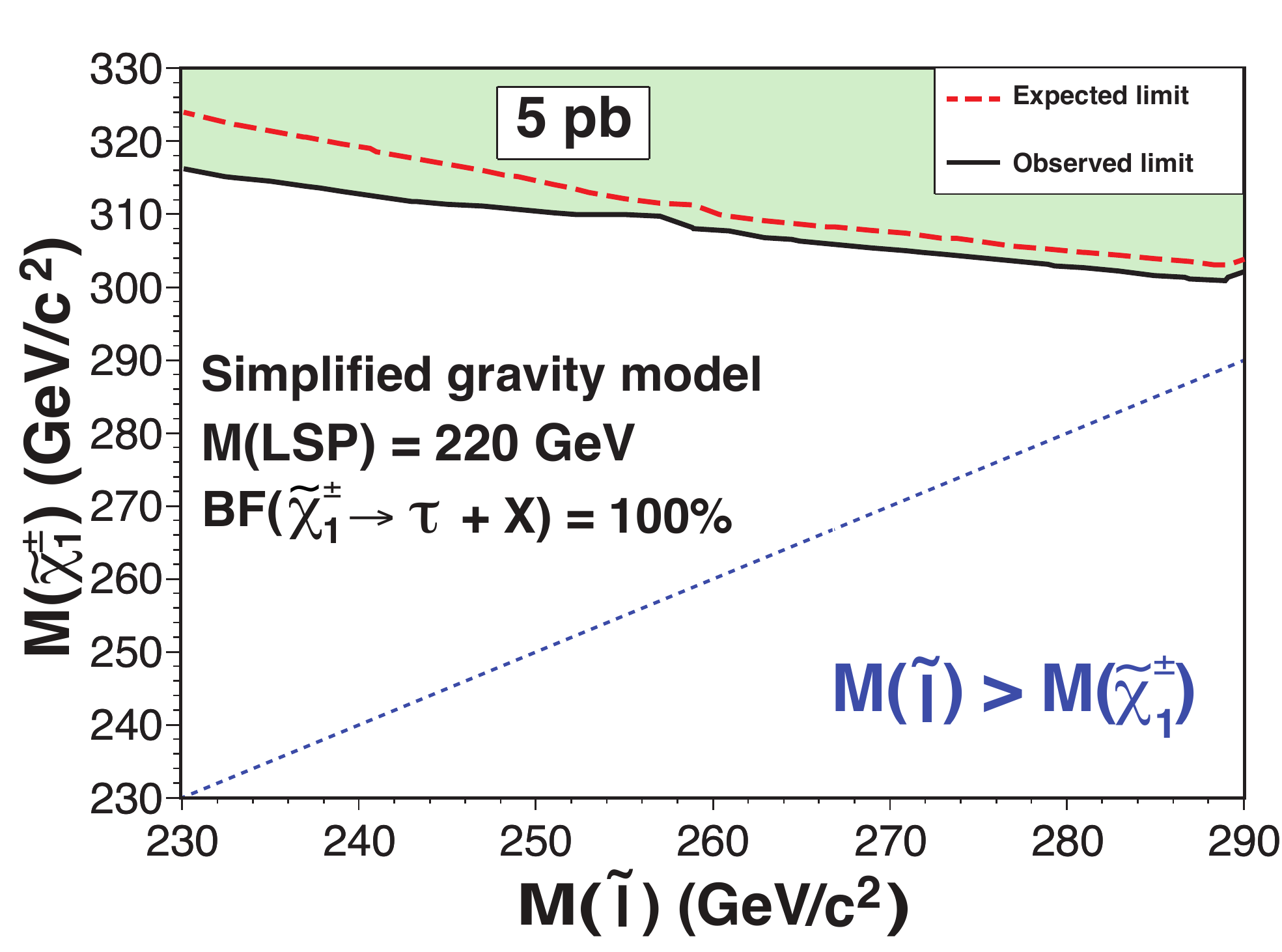}
    \caption{Expected and observed  contours of constant  95\% C.L. cross-section upper limits in the chargino-slepton mass plane assuming the simplified gravity-mediated model for ${\it BF}(\tilde{\chi} \to \tau + X)=33\%$ and $100\%$, for two different values of LSP mass.  The shaded regions correspond to cross section limits of $\sigma(\cone\ntwo) \leq $ 5 pb, as functions of the gaugino and slepton masses.
    \label{fig:limits-gravity}}
  \end{center}
\end{figure}

Given the good agreement between the data and the background prediction, we
interpret the results as exclusion limits on the rates of SUSY processes.  We set upper limits at 95\% credibility level (C.L.) on the cross section for chargino-neutralino production as a
function of chargino mass (assumed mass degenerate with \ntwo), slepton mass, LSP mass (for the case of the simplified gravity-mediated model), 
and branching fraction of the chargino (and neutralino) to the stau.  
Limits are extracted using a Bayesian technique and incorporating the systematic uncertainties described
above \cite{mclimit}.  We generate SUSY signal samples using \textsc{madgraph} \cite{Maltoni:2002qb}.  For each set of signal parameters we optimize the \met\ requirement above 20 \gev\ to minimize the median value of the excluded cross section assuming the observation exactly matches the background prediction (expected limit).  The chosen value accounts for the various differences between the SUSY particle masses, while the 20 \gev\ minimum value is motivated by the selection in Ref.~\cite{Abulencia:2007rd}.  Table~\ref{tab:bgdata} also shows a comparison of an example signal with the background expectation and data before and after this requirement.
Representative cross-section upper limit contours are
shown in Figs.~\ref{fig:limits-gauge} and~\ref{fig:limits-gravity} for simplified gauge- and gravity-mediated models.  We emulate the effect of raising $\tan\beta$ by directly altering the branching fraction of the chargino and neutralino to a stau, and consider both 33\% and 100\%, corresponding to lepton universality and tau-dominated scenarios, respectively.  For the simplified gravity-mediated model, we determine limit contours for $m(\tilde{\chi}^0_1) = 120$ and $220 \gevcc$. 
As the chargino and neutralino masses increase, the cross-section limits for both models become more stringent due to the increased acceptance, and then vanish at the Tevatron kinematic limit for new particle production, corresponding to 1.96 TeV for the mass sum for all produced particles.
The gaps in exclusion at high mass between the exclusion curves and the kinematic limits, shown as diagonal lines, are due to the tau and lepton \pt\ requirements as well as the optimized \met\ requirements
for each mass pair.

In summary, we search for a like-sign lepton-tau signal in CDF Run II data corresponding to 6.0 fb$^{-1}$ of integrated luminosity.  This distinctive signature is expected to be 
sensitive to SUSY models with direct chargino-neutralino associated production.  Observing no
significant excess of events in the data over standard model background predictions, we set upper limits on the cross section
for this SUSY process as a function of the sparticle masses and branching fractions to taus.
Our results, presented in simplified gravity- and gauge-mediated frameworks, are complementary to SUSY searches that require substantial hadronic jet activity.  This analysis also constrains regions of electroweak gaugino production at high $\tan\beta$, where decays to taus dominate, and gauge-mediated parameter space with slepton next-to-lightest SUSY particles for the first time.

We thank Howie Baer, Markus Luty, Natalia Toro, Josh Ruderman, and David Hsih for
theoretical guidance.
We thank the Fermilab staff and the technical staffs of the participating institutions for their vital contributions. This work was supported by the U.S. Department of Energy and National Science Foundation; the Italian Istituto Nazionale di Fisica Nucleare; the Ministry of Education, Culture, Sports, Science and Technology of Japan; the Natural Sciences and Engineering Research Council of Canada; the National Science Council of the Republic of China; the Swiss National Science Foundation; the A.P. Sloan Foundation; the Bundesministerium f\"ur Bildung und Forschung, Germany; the Korean World Class University Program, the National Research Foundation of Korea; the Science and Technology Facilities Council and the Royal Society, UK; the Russian Foundation for Basic Research; the Ministerio de Ciencia e Innovaci\'{o}n, and Programa Consolider-Ingenio 2010, Spain; the Slovak R\&D Agency; the Academy of Finland; the Australian Research Council (ARC); and the EU community Marie Curie Fellowship contract 302103.

\bibliography{LStau-prl}

\begin{thebibliography}{41}
\bibitem [{cdf()}]{cdfgeometry}%
  \BibitemOpen
  \href@noop {} {}\bibinfo {note} {CDF uses a right-handed cylindrical
  coordinate system with the origin at the center of the detector, the $z$ axis
  in the direction of the proton beam, and $\theta$ and $\phi$ denoting the
  polar and azimuthal angles, respectively. Pseudorapidity is defined $\eta = -
  \ln \tan \theta/2$. The transverse momentum and energy of a particle or jet
  are defined as $p_T = p\sin\theta$ and $E_T= E \sin\theta$.}\BibitemShut
  {Stop}%
\end{thebibliography}
\end{document}